\documentclass[12pt]{amsart}
\usepackage{amssymb}
\usepackage{verbatim}
\usepackage[usenames]{color}
\usepackage{soul}
\setstcolor{green}
\usepackage{hyperref}
\usepackage{url}

\newtheorem{thm}{Theorem}

\theoremstyle{definition}

\theoremstyle{remark}
\newtheorem{ex}[thm]{Example}
\newtheorem{rmk}[thm]{Remark}

\newcommand{\bbb}[1]{\ensuremath{\mathbb {#1}}}
\newcommand{\ttt}[1]{\ensuremath{\mathtt {#1}}}

\newcommand{\emp}{\varnothing}
\renewcommand{\phi}{\varphi}

\newcommand{\sq}[1]{\ensuremath{\langle#1\rangle}}

\newcommand{\notarrow}{\kern .42em\not\kern -.42em\longrightarrow}

\newenvironment{eatab}
{\bigskip\noindent\begin{minipage}{\textwidth}\upshape\ttfamily
\begin{tabbing}mmm\=mmm\=mmm\=mmm\=mmm\=\kill}
{\end{tabbing}\end{minipage}\bigskip}

\newcommand{\noprint}[1]{\relax}

\title{When Are Two Algorithms the Same?}
\author{Andreas Blass}
\address{Mathematics Department\\
  University of Michigan\\
  Ann Arbor, MI 48109--1043, U.S.A.}
\email{ablass@umich.edu}
\thanks{Blass was partially supported by NSF grant DMS-0653696 and a
  grant from Microsoft Research.  Dershowitz's research was supported
  in part by the Israel Science Foundation under grant no.\ 250/05.
  Part of the work reported here was done at Microsoft Research.}
\author{Nachum Dershowitz}
\address{School of Computer Science, Tel
  Aviv University, 69978 Ramat Aviv, Israel}
\email{nachum.dershowitz@cs.tau.ac.il}
\author{Yuri Gurevich}
\address{Microsoft Research\\
  One Microsoft Way\\
  Redmond, WA \ 98052, U.S.A.}
\email{gurevich@microsoft.com}

\begin{document}

\begin{abstract}
  People usually regard algorithms as more abstract than the programs
  that implement them.  The natural way to formalize this idea is that
  algorithms are equivalence classes of programs with respect to a
  suitable equivalence relation.  We argue that no such equivalence
  relation exists.
\end{abstract}

\maketitle

\section{Introduction}

At the end of his contribution to a panel discussion on logic for the
twenty-first century \cite[page~175]{shore}, Richard Shore posed, as the
last of three ``probably pie-in-the-sky'' problems, the following:

\begin{quote}
  Find, and argue conclusively for, a formal definition of algorithm
  and the appropriate analog of the Church-Turing thesis. Here we want
  to capture the intuitive notion that, for example, two particular
  programs in perhaps different languages express the same algorithm,
  while other ones that compute the same function represent
  different algorithms for the function. Thus we want a definition
  that will up to some precise equivalence relation capture the notion
  that two algorithms are the same as opposed to just computing the
  same function.
\end{quote}

The purpose of this paper is to support Shore's ``pie-in-the-sky''
assessment of this problem by arguing that there is no satisfactory
definition of the sort that he wanted.  That is, one cannot give a
precise equivalence relation capturing the intuitive notion of ``the
same algorithm.''  We describe several difficulties standing in
the way of any attempt to produce such a definition, and we give
examples indicating that the intuitive notion is not sufficiently
well-defined to be captured by a precise equivalence relation.

Following the terminology in Shore's statement of the problem, we use
``program'' for an element of the set on which the desired equivalence
relation is to be defined and we use ``algorithm'' for an equivalence
class of programs (or for the abstract entity captured by an
equivalence class) with respect to an equivalence relation of the sort
requested by Shore.

Shore's question has been addressed in a limited context by Yanofsky
\cite{yanof}.  He considers programs only in the form of (fairly)
standard constructions of primitive recursive functions, and he
defines an equivalence relation on these programs by listing a number
of situations in which two programs should be considered equivalent.
He does not, however, claim that the equivalence relation generated by
these situations completely captures the intuitive notion of
equivalence of programs.  In fact, he explicitly says ``it is doubtful
that we are complete,'' and he anticipates one of our objections to
the ``equivalence class of programs'' view of algorithms by saying
that ``whether or not two programs are the same \dots\ is really a
subjective decision.''  By considering only a limited class of
programs and limited sorts of equivalence between them, Yanofsky
obtains a notion of algorithm with pleasant properties from the point
of view of category theory, but not what Shore asked for.

We shall also consider only a limited class of algorithms, namely
deterministic, small-step, non-interactive algorithms.  But for our
purposes, such a limitation strengthens our argument.  If no suitable
equivalence relation can be found in this limited case, then the
situation is all the more hopeless when more algorithms (distributed
ones, interactive ones, etc.) enter the picture.  Any suitable
equivalence relation on a large class of programs would restrict to a
suitable equivalence relation on any subclass.

Small-step algorithms are those that proceed in a discrete sequence of
steps, performing only a bounded amount of work per step.  (The bound
depends on the algorithm but not on the input or on the step in the
computation.)  Non-interactive algorithms\footnote{It is argued in
  \cite[Section~9]{seqth} that non-determinism is a special case of
  interaction.  Thus, ``non-interactive'' implicitly entails
  ``deterministic.''} of this sort were characterized axiomatically in
\cite{seqth}.  They include traditional models of computation such as
Turing machines and register machines.  Most people, when they hear
the word ``algorithm'', think of deterministic, small-step,
non-interactive algorithms unless a wider meaning is explicitly
suggested.  And until the 1960's, essentially all algorithms in the
literature were of this sort.  Because Shore posed his question in the
context of traditional recursion theory, it is reasonable to suppose
that he had deterministic, small-step, non-interactive algorithms in
mind.  If so, then it is not only reasonable but in a sense necessary
to argue our case in the limited context of such algorithms.  For if
an argument against the existence of the desired equivalence relation
depended on other sorts of algorithms (for example, interactive ones
or massively parallel ones), then that argument might be regarded as
missing the point of Shore's question.

In Sections~2 and 3, we discuss preliminary issues concerning the
choice of programming languages and their semantics.  In Section~4, we
describe situations where people can reasonably disagree as to whether
two programs implement the same algorithm.  In Section~5, we discuss
how the intended use of algorithms can affect whether they should
count as the same.  Section~6 contains examples where two quite
different algorithms are connected by a sequence of small
modifications, so small that one cannot easily say where the
difference arises.  Finally, in Section~7, we mention several other
domains in which analogous issues arise.

\section{Programs}

In order to even hope to capture the notion of algorithm by a precise
equivalence relation on the set of programs, it is necessary first to
have a precise set of programs to work with.  The present section
addresses this preliminary issue.  We argue that there is a precise
set of programs, at least for small-step, non-interactive algorithms,
but that this fact is not as obvious as one might think at first.

The mention, in Shore's question, of ``programs in perhaps different
languages'' suggests that the set of programs should be rather wide,
encompassing programs written in a variety of languages.  But caution
is needed here, in order to keep the set of programs well-defined.  We
cannot allow arbitrary programming languages, because new ones are
being created (and perhaps old ones are falling out of use) all the
time. Even languages that persist over time tend to gradually acquire new
dialects.  We should work instead with a specific set of stable
languages.

Not only should the languages be stable, so as not to present a moving
target for an equivalence relation, but they should have a precise,
unambiguous semantics.  For example, in some languages, such as C, the
order of evaluation of subexpressions in an expression may be left to
the discretion of the compiler, and whether this ambiguity matters in
a particular program is, in general, undecidable.  If deterministic
algorithms are to be captured by equivalence classes of programs, then
ambiguity in the programs is unacceptable.  This consideration
excludes a lot of languages.  And pseudo-code, despite being widely
used to describe algorithms, is even worse from the point of view of
precision.

For many purposes in theoretical computer science, especially in
complexity theory, it is traditional to use, as the official
framework, low-level models of computation, such as Turing machines or
register machines, that have an unambiguous semantics.  Such models of
computation are, however, inadequate to serve as the domain of an
equivalence relation capturing the general notion of algorithm.  The
problem is that they can express algorithms only at a low level of
abstraction, with many implementation details that are irrelevant to
the algorithm.

At first sight, it seems that this difficulty can be circumvented by a
suitable choice of equivalence relation on, say, Turing machine
programs.  Programs that differ only in the irrelevant details of
implementation should be declared equivalent.  There remains, however,
the problem that, by including such details, a Turing machine program
may make the intended algorithm unrecognizable.  It is not in general
easy to reverse engineer a Turing machine program and figure out what
algorithm it implements.  In fact, it may be impossible; here is an
example illustrating the problem (as well as some other issues to be
discussed later).

\begin{ex} \label{euclid1} The Euclidean algorithm for computing the
  greatest common divisor (g.c.d.) of two positive integers is
  described by the following abstract state machine (ASM), but readers
  unfamiliar with ASM notation should be able to understand it as if
  it were pseudo-code.  In accordance with the usual conventions
  \cite{lipari} for ASMs, the program is to be executed repeatedly
  until a step leaves the state unchanged.  We assume for simplicity
  that the numbers whose g.c.d.\ is sought are the values, in the
  initial state, of dynamic nullary function symbols $x$ and $y$.  We
  use \ttt{rem} for the remainder function; $\ttt{rem}(a,b)$ is the
  remainder when $a$ is divided by $b$.

\begin{eatab}
  if rem$(y,x)=0$ \\
then output$:=x$\\
else do in parallel $x:=$rem$(y,x)$, $y:=x$ enddo \\
endif
\end{eatab}

Here is a variant, using subtraction instead of division.  In fact,
this variant corresponds more closely to what Euclid actually wrote;
see \cite[Propositions~VII.2 and X.3]{joyce}.

\begin{eatab}
if $x<y$ then $y:=y-x$\\
elseif $y<x$ then $x:=x-y$\\
else output$:=x$\\
endif
\end{eatab}

Notice that the second version in effect uses repeated subtraction
(rather than division) to compute in several steps the remainder
function \ttt{rem} used in a single step of the first version.

Are these two versions the same algorithm?  As we shall discuss later,
this issue is debatable.  The point we wish to make here, before
entering any such debate, is that the use of low-level programs like
Turing machines might cut off the debate prematurely by making the two
versions of Euclid's algorithm identical.  Consider programming these
versions of Euclid's algorithm to run on a Turing machine, the input
and output being in unary notation.  The most natural way to implement
division in this context would be repeated subtraction.  With this
implementation of division, any difference between the two versions
disappears.

If we were to use Turing machines (or similar low-level programs) as
our standard, then to avoid letting them preempt any discussion of
whether the two versions of Euclid's algorithm are the same, we would
have to implement at least one of the two versions in a special way.
For example, we might implement division by converting the inputs to
decimal notation and having the Turing machine perform division as we
learned to do it in elementary school.  But then the question arises
whether the preprocessing from unary to decimal notation has altered
the algorithm.  And once preprocessing enters the picture, there are
arbitrary choices to be made (decimal or binary?), which further
complicate the picture.
\end{ex}

Motivated by examples like the preceding one, we refuse to restrict
ourselves to low-level programming languages.  We want a language (or
several languages) capable of expressing algorithms at their natural
level of abstraction, not forcing the programmer to decide
implementation details.  This desideratum, combined with the earlier
one that the language(s) should have a precise, unambiguous semantics,
severely restricts the supply of candidate languages.  In fact, the
only such language that we are aware of is the abstract state machine
language, which, in our present context, means small-step (also called
sequential) ASMs, as presented in \cite{seqth}.

\begin{rmk}
  To make this paper more self-contained, we give the following brief
  description of sequential ASMs.  For more details, see \cite{ch-th}
  or \cite{lipari} or \cite{seqth}.  The rest of this paper does not
  depend on the exact definition of ASMs; a reader willing to read ASM
  programs as pseudo-code should have no difficulty with their use
  here.

An ASM program describes operations to be performed on a state.  The
state is a structure in the usual sense of first-order logic.  The
computation progresses in discrete steps, changing the interpretations
of certain function symbols in the state.  The basic units from which
ASM programs are built are \emph{update rules} of the form
$f(t_1,\dots,t_k):=t_0$, whose meaning is that the value of (the
interpretation of) the $k$-ary function symbol $f$ at the $k$-tuple
consisting of (the values of) $t_1,\dots,t_k$ is to be changed to (the
value of) $t_0$.  ASM programs are built from update rules by two
constructors.  One is parallel composition;
\[
\ttt{do\ in\ parallel\ }R_1,\dots,R_n\ttt{\ enddo}
\]
means to make all the state changes given by the rules
$R_1,\dots,R_n$.  (If two of those changes clash, prescribing
different updates of the same function at the same arguments, then no
changes are to be made.)  The other constructor is the conditional
\[
\ttt{if\ }\phi\ttt{\ then\ }R_0\ttt{\ else\ }R_1,
\]
with the obvious meaning; here $\phi$ is a quantifier-free first-order formula.  This
concludes our rough description of ASM programs.
\end{rmk}

We note here another advantage of ASMs: They can deal directly with
structures, such as graphs, as input and output.  This contrasts with
the need, in most other programming languages, to code structures.  In
the case of graphs, coding would involve not only deciding, for
example, to represent graphs as adjacency matrices, but also choosing,
for each particular input graph, an ordering of its vertices, because
an adjacency matrix presupposes a particular ordering.

ASMs also make it possible to write programs that use powerful,
high-level operations.  This is important for our purposes because we
do not want the programming language to prejudge whether an algorithm
using such operations is equivalent to one that implements those
operations by means of lower-level ones. To avoid prejudging, the
language must be able to formulate separately the high-level and
low-level versions, not simply use the latter as a surrogate for the
former.  The possibility of writing programs at different levels is
one of the key strengths of ASMs.

In view of these considerations, we shall use ASMs as our standard
programs in this paper.  We emphasize, however, that this decision is
not essential for most of our arguments.  The difficulties facing an
attempt to define the ``right'' equivalence relation on programs are
intrinsic difficulties, not caused by our decision to use ASMs.  In
fact, the difficulties would be even worse if we were to use
programming
languages that lack a precise semantics or programming languages that
cannot work at multiple levels of abstraction.

\section{Behavioral Equivalence}  \label{beh-eq}

A proper understanding of the use of ASMs to describe algorithms
requires a look at the concept of \emph{behavioral equivalence}, which
plays a central role in the theory of ASMs.  It is based on an
abstract view of algorithms presented in \cite{seqth}. Before
discussing this point of view, it is necessary to resolve a
terminological ambiguity, because the word ``algorithm'', which plays
a crucial role in Shore's question, is also used as an abbreviation
for ``sequential algorithm'' (as defined in \cite{seqth}) when no
other sorts of algorithms are under consideration.  We shall be
careful to abstain from this abbreviation.  ``Sequential algorithm''
will mean what is defined in \cite{seqth}, namely an entity to which
are associated a set of states, a set of initial states, and a
transition function, subject to certain postulates;\footnote{A rough
  summary of the postulates is that states are first-order structures,
  the transition function involves only a bounded part of the state,
  and everything is invariant under isomorphisms of structures.}
``algorithm'' without ``sequential'' will mean what Shore asked for
(and what we claim admits no precise characterization), essentially an
equivalence class of programs with respect to the pie-in-the-sky
equivalence relation.  The terminology is somewhat unfortunate in that
(1)~it suggests that ``sequential algorithms'' are a special case of
``algorithms'', whereas the two concepts are actually separate and
(2)~the word ``sequential'' that is the difference between the two
names is not a difference between the concepts, since we retain our
earlier convention that the only algorithms (in Shore's sense) that we
consider here are small-step, non-interactive ones.  Nevertheless, it
seems reasonable to use the expressions ``sequential algorithm'' and
``algorithm'' to match the terminology of the sources, \cite{seqth}
and \cite{shore} respectively.

Behavioral equivalence, defined in \cite{seqth} (but there called
simply ``equivalence'') is a very fine equivalence relation on
sequential algorithms, requiring that they have the same states, the
same initial states, and the same transition function.  The main
theorem in \cite{seqth} is that every sequential algorithm is
behaviorally equivalent to a (sequential) ASM.  As was pointed out in
\cite{seqth}, behavioral equivalence seems excessively fine for many
purposes, but this circumstance only makes the main theorem of
\cite{seqth} stronger.

For our present purposes, it is useful, though not absolutely
necessary, to know that behavioral equivalence is fine enough, that
is, that two behaviorally equivalent sequential algorithms are
equivalent in Shore's sense.  This knowledge is useful because it
ensures that, by using ASMs as our standard programs, we have
representatives for all of the algorithms in Shore's sense.  If it
were possible for behaviorally equivalent programs to be inequivalent
in Shore's sense, then there might be algorithms that, although
behaviorally equivalent to ASMs, are not Shore-equivalent to any ASMs,
and such algorithms would be overlooked in our discussion.  This would
not be a disaster -- if we convince the reader that there is no precise definition
for Shore-equivalence on ASMs, then there is surely no such definition
for an even wider class of programs (this is why we described the
knowledge above as not absolutely necessary)---but it is desirable to
know that we are dealing with the whole class of algorithms that we
intended to deal with.

Is this desirable knowledge in fact correct?  Must behaviorally
equivalent algorithms be equivalent in the sense that Shore asked for?
In the rest of this section, we argue that the answer is yes, as long
as we do not read into ASMs more information than they are intended to
convey.  In brief, the point is that ASMs are intended to describe
sequential algorithms and, since these are not specified beyond their
states, initial states, and a transition function,\footnote{The
  definition of ``sequential algorithm'' \cite{seqth} says that they
  are entities associated with the three items mentioned, but the
  postulates concern only these three items.  No further role is
  played by the entities themselves.} there is no room for any
distinction between behaviorally equivalent ASMs or sequential
algorithms.

Because the ASM syntax is readable as pseudo-code, however, there is a
temptation to consider an ASM as providing additional information
beyond the sequential algorithm that it defines.  Here is an example
to illustrate the situation.

\begin{ex} \label{dec-tree} We describe here, in ASM form, a binary
  decision tree algorithm.  The vocabulary has 0-ary relation symbols
  (also known as propositional symbols) $U_s$ for every sequence
  $s\in\{0,1\}^{\leq5}$ (i.e., every sequence of at most 5 binary
  digits), it has constant symbols $c_s$ for $s\in\{0,1\}^6$, and it
  has one additional 0-ary symbol \ttt{output}.  (Visualize the
  symbols $U_s$ as attached to the internal nodes $s$ of a binary tree
  of height 7 and the symbols $c_s$ as attached to the leaves.)  The
  only dynamic symbol is \ttt{output}.  States are arbitrary
  structures for this vocabulary.  The ASM program we want is best
  described by reverse induction on binary strings $s$ of length
  $\leq6$.  If $s$ has length 6, let $\Pi_s$ be the program
\[
\ttt{output}:=c_s.
\]
If $s$ has length $\leq5$, let $\Pi_s$ be
\[
\ttt{if}\ U_s\ \ttt{then}\ \Pi_{s^\frown\sq1}\ \ttt{else}\
\Pi_{s^\frown\sq0}.
\]
The program we are interested in is $\Pi_{\emp}$ where $\emp$ is the
empty sequence.  Its natural, intuitive interpretation is as a
computation that begins by looking up $U_\emp$; depending on the
answer, it looks up $U_s$ for $s=\sq0$ or $s=\sq1$; depending on the
answer, it looks up $U_s$ for an appropriate $s$ of length 2; and so
forth, until, after 6 queries, it sets \ttt{output} to the appropriate
$c_s$.  In other words, we have a standard binary decision tree.

The algorithm does all this work in one step.  (If it were allowed to
run for another step, nothing would change in the state.)  Notice that
the program has length of the order of magnitude $2^6$ and that its
natural interpretation as a computational process looks up 7 values,
namely 6 $U_s$'s and one $c_s$.  (To make precise sense of ``order of
magnitude'' here and similar notions below, pretend that 6 is a
variable.)

Now let us apply the proof of the main theorem in \cite{seqth} to the
sequential algorithm defined by this ASM.  The proof produces another
ASM, $\Pi'$, behaviorally equivalent to our $\Pi_\emp$.  This ASM (see
\cite[Lemma~6.11]{seqth}) is a parallel combination of guarded
updates, one for each of the $2^{63}$ possible systems of values of
the $U_s$'s.  So $\Pi'$ has length of the order of magnitude
$2^{2^6}$, exponentially greater than $\Pi_\emp$.  Furthermore, in
each run of the natural, intuitive interpretation of $\Pi'$ as a
computational process, it looks up the values of all 63 of the
$U_s$'s.
\end{ex}

In view of the vastly greater work done by $\Pi'$ in each run (and its
vastly greater length), it would be difficult to convince people that
it represents the same algorithm as $\Pi_\emp$.  Yet, all the
complexity of $\Pi'$ is hidden if one looks only at states and
transition functions, and so $\Pi'$ is behaviorally equivalent to
$\Pi_\emp$.  We assert that $\Pi'$ and $\Pi_\emp$ should be regarded
as defining the same algorithm, that is, as equivalent programs in
Shore's sense.  The apparent differences between them arise from
regarding the ASM programs not merely as descriptions of sequential
algorithms but as telling in detail how the transition function is to
be computed.  It is tempting to assign this sort of detailed
operational semantics to ASMs, but this conflicts with our intention
to use them, as in \cite{seqth}, simply to describe sequential
algorithms.

We do not claim that one should ignore the difference between an
algorithm that looks up seven values and performs six tests on them
and an algorithm that looks up 63 values and performs approximately
$2^{63}$ tests on them.  But we do claim that, if these properties of
the algorithms are to be taken into account, then the ASMs
representing the algorithms should be designed to make these
properties visible in the states and transition functions.  That this
can be done is part of the ASM thesis: Any algorithm can be
represented at its natural level of abstraction by an ASM.  In the
case at hand, the level of abstraction could involve keeping track of
which $U_s$'s the algorithm has used.
This can be done, for example, by having additional dynamic symbols
$V_s$ (with value \ttt{false} or \ttt{undef} in initial states) with
the intended meaning ``$U_s$ has been used'' and by including in both
$\Pi_\emp$ and $\Pi'$ updates setting $V_s$ to \ttt{true} whenever
$U_s$ is used.\footnote{Another way to make the evaluation of $U_s$'s
  visible for the purposes of behavioral equivalence is to make the
  $U_s$'s external function symbols, so that evaluating them involves
  a query to the environment.  In this paper, however, we prefer to
  consider only non-interactive algorithms.}  Similarly, if we wish
to take into account the huge number of guards (of conditional rules)
evaluated in the computation intuitively described by $\Pi'$, then we
should write the ASM at that level of abstraction, that is, we should
include updates whereby the algorithm keeps a record of all this work.

For the rest of this paper, we shall take for granted, with the
support of considerations like the preceding paragraph, that, whatever
Shore's pie-in-the-sky equivalence relation ought to be, it will be at
least as coarse as behavioral equivalence.

\begin{rmk}
  Udi Boker [personal communication] has suggested that, once it is
  clear that ASMs are to serve as descriptions of what is to happen as
  a result of a single step, not caring about how it is accomplished
  within the step, then behavioral equivalence captures the intuitive
  notion of algorithm. We feel that, although behavioral equivalence
  is a natural and important equivalence relation, it need not capture
  the intution.  In Example~\ref{euclid1} above and the related
  Example~\ref{euclid2} below, we have algorithms that are not
  behaviorally equivalent but can nevertheless be reasonably viewed as
  the same algorithm.  In other words, the intuitive notion of ``the
  same algorithm'' seems not to require the computations to match step
  by step.  Nor does it require the vocabulary of the states to be the
  same; a reasonable renaming of identifiers in a program should not
  change the algorithm (but see also Example~\ref{rename} below).
\end{rmk}

Given that Shore's pie-in-the-sky equivalence relation should be at
least as coarse as behavioral equivalence, the main theorem of
\cite{seqth} assures us that ASMs are adequate for representing all
sequential algorithms in the sense of \cite{seqth}.  These are exactly
the small-step, non-interactive algorithms that we intend to treat in
this paper.  We therefore have an adequate set of programs, the ASMs,
with a precisely defined semantics.  That is, we have a set of
programs on which it makes sense to try to define an equivalence
relation of the sort Shore asked for.  Now that this prerequisite for
Shore's question is satisfied, we turn to the question itself.

\section{Subjectivity}

The formulation of Shore's question, to ``capture the notion that two
algorithms are the same,'' presupposes that there is such a notion.
And indeed, people do seem to have such a notion, to use it, and to
understand each other when they use it.  But it is not clear that they
all use it the same way.  Will two competent computer scientists (or
mathematicians or programmers) always agree whether two algorithms are
the same?  We have already quoted Yanofsky's doubt about this point:
equivalence of algorithms is a subjective notion.

There are situations where disagreement is almost guaranteed.  Suppose
X has invented an algorithm and Y later invents a somewhat modified
version of it. X will be more likely than Y to say that the algorithms
are really the same.

Even when ulterior motives are not involved, there can easily be
disagreements about whether two algorithms are the same.  Consider,
for example, the two versions of the Euclidean algorithm in
Example~\ref{euclid1}, one using division and the other using
repeated subtraction.  Are they the same?

One can argue that they are different.  The version with subtraction
is slower.  It can be implemented on simpler processors, for instance,
children who have learned how to subtract but not how to divide.  It
is not behaviorally equivalent to the division version because it
takes several steps to do what division did in one step.

Despite all these differences, it seems that most mathematicians (and
computer scientists?) would, when confronted with either of these
algorithms, call it the Euclidean algorithm.  Furthermore, if one just
asks, ``What is the Euclidean algorithm?'', the answer is usually the
division version (see for example \cite[Section~8.8]{ahu} or
\cite{wiki}), even though Euclid gave the subtraction version.  To
call the division version ``Euclidean'' strongly suggests that it is
considered ``the same'' as what Euclid did.

The subjectivity of the situation is further emphasized by the
following variant.

\begin{ex} \label{euclid2}
  For positive integers $a$ and $b$, let $\ttt{app}(a,b)$ be the
  multiple of $b$ that is closest to $a$ (possibly below $a$ and
  possibly above, and in case of a tie then below), and let
\[
\ttt{rem}'(a,b)=|a-\ttt{app}(a,b)|
\]
be the distance from $a$ to the nearest multiple of $b$.  Now replace
\ttt{rem} by $\ttt{rem}'$ in the division version of the Euclidean
algorithm.  Does that change the algorithm?  Notice that it may
happen that, in runs of the \ttt{rem} and $\ttt{rem}'$ versions of the
algorithm on the same input, only the initial and final states
coincide; all the intermediate states may be different (and there may
be different numbers of intermediate states).  Nevertheless, both
versions are called the Euclidean algorithm.
\end{ex}

Finally, what about the Euclidean algorithm applied not to integers
but to polynomials in one variable (over a field)?  It proceeds just
like the division version in Example~\ref{euclid1}, but it uses
division of polynomials.  When $a$ and $b\neq0$ are polynomials, there
are unique polynomials $q$ and $r$ such that $a=bq+r$ and $r$ has
strictly smaller degree than $b$ does.  (We regard the zero polynomial
as having degree $-\infty$.)  Let $\ttt{Rem}(a,b)$ denote this $r$,
and replace \ttt{rem} by \ttt{Rem} in the division version above to
get the Euclidean algorithm for polynomials.  Is that a different
algorithm?

\begin{rmk}   \label{ideal}
There is a more abstract, non-algorithmic proof that every two
positive integers have a g.c.d.  It consists of showing that the
smallest positive element $z$ of the ideal $\{mx+ny:m,n\in\bbb Z\}$
serves as the g.c.d.  The main point of the proof is to show that this
$z$ divides both $x$ and $y$.  That proof implicitly contains an
algorithm for finding the g.c.d.\ $z$.  As before, we take the initial
state to have $x$ and $y$, the two numbers whose g.c.d.\ is sought,
but this time $x$ and $y$ are static; there is one dynamic function
symbol $z$, initially equal to $x$.

\begin{eatab}
if $\ttt{rem}(x,z)\neq0$ then $z:=\ttt{rem}(x,z)$\\
elseif $\ttt{rem}(y,z)\neq0$ then $z:=\ttt{rem}(y,z)$\\
else output$:=z$\\
endif
\end{eatab}

It seems that this algorithm is really different from the Euclidean
one; it uses a different idea.  But we are not prepared to guarantee
that no one will regard it as just another form of the Euclidean
algorithm.  Nor do we guarantee that ``the same idea'' has a clear
meaning.
\end{rmk}

Here are some more examples where it is debatable whether two
algorithms should be considered the same.

\begin{ex}   \label{format}
  The first is actually a whole class of examples.  Consider two
  algorithms that do the same calculations but then format the outputs
  differently.  Are they the same algorithm?  Our description of the
  situation, ``the same calculations'' suggests an affirmative answer,
  but what if the formatting involves considerable work?
\end{ex}

\begin{ex}
  Is a sorting algorithm, like quicksort or mergesort a single
  algorithm, or is it a different algorithm for each domain and each
  linear ordering?  This question is similar to the earlier one about
  the Euclidean algorithm for numbers and for polynomials, and it is
  clear that one could generate many more examples of the same
  underlying question.
\end{ex}

Notice that there are some situations that can be regarded as falling
under both of the preceding two examples.  Suppose one algorithm sorts
an array of numbers into increasing order and a second algorithm sorts
the same inputs into decreasing order.  They can be viewed as the same
algorithm with different formatting of the output, or they can be
viewed as the same algorithm applied to two different orderings of the
natural numbers.

\begin{ex} \label{pivots} Quicksort begins by choosing a pivot element
  from the array to be sorted.  One might always choose the first
  element, but a better approach is to choose an element at random.  A
  variant uses the median of three elements (for example, the first,
  middle, and last entries of the array) as the pivot.  (See, for
  example, \cite[pages~94 and 95]{ahu}.)
  Are these different
  algorithms? What if one increases ``three'' to a larger odd number?
\end{ex}

\section{Purpose}

In the preceding section, we attributed variations in the notion of
``the same algorithm'' to people's different opinions and subjective
judgments.  There is, however, a more important source of variation,
even between the notions used by the same person on different
occasions, namely the intended purpose.

For someone who will implement and run the algorithm, differences in
running time are likely to be important.  A difference in storage
requirements may be very important on a small device but not on a
device with plenty of space.  Of course, the distinction between a
small device and one with plenty of space depends on the nature of the
computation; even a supercomputer is a small device when one wants to
predict next week's weather.  Thus, when one considers a specific
device and asks whether an algorithm's space requirements are an
essential characteristic of the algorithm, that is, whether one should
count two algorithms as different just because one uses much more
space than the other, then answer is likely to be ``yes'' once the
space requirements are large enough but ``no'' if they are small.
Thus, two programs may well count as the same algorithm for someone
programming full-size computers but not for someone programming
cell-phones.

Although we generally consider only non-interactive algorithms in this
paper, let us digress for one paragraph to consider (a limited degree
of) interaction.  If an algorithm is to be used as a component in a
larger system, additional aspects become important, which may be
ignored in stand-alone algorithms.  And in this respect, even just the
interaction between an algorithm and the operating system can be
important for some purposes.  Two programs that do essentially the
same internal work but ask for different allocations of memory space
(or ask for the same allocation at different stages of the
computation) or that make different use of library calls will be
usefully considered as different algorithms for some purposes but
probably not for all.  Similarly, if an algorithm's output is to be
used as the input of some other calculation, then formatting (see
Example~\ref{format}) is more important than it otherwise would be.

Returning to the standard situation of non-interactive, small-step
algorithms, we present an example which, although it arose in a
different context (complexity theory), helps to show how one's purpose
can influence whether one considers two algorithms the same.

\begin{ex}
  Suppose we have programs $P$ and $Q$, computing functions $f$ and
  $g$, respectively, say from strings to strings, and each requiring
  only logarithmic space and linear time (in a reasonable model of
  computation).  Can we combine $P$ and $Q$ to compute the composite
  function $f\circ g$ in linear time and log space?  The simplest way
  to compute $f\circ g$ on input $x$ would be to first use $Q$ on
  input $x$ to compute $g(x)$ and then use $P$ on input $g(x)$ to
  compute $f(g(x))$.  This combined algorithm runs in linear time but
  not in log space; storing the intermediate result $g(x)$ will in
  general require linear space.

There is an alternative method to compute $f(g(x))$.  Begin by running
$P$, intending to use $g(x)$ as its input but not actually computing
$g(x)$ beforehand.  Whenever $P$ tries to read a character in the
string $g(x)$, run $Q$ on input $x$ to compute that character and
provide it to $P$.  But make no attempt to remember that computation
of $Q$.  When $P$ needs another character from $g(x)$, run $Q$ again
to provide it.  Because this version of the algorithm does not try to
remember $g(x)$, it can run in log space.  But because it needs to
restart $Q$ every time $P$ wants a character from $g(x)$, it will run
in general in quadratic rather than linear time.

So we have two programs that combine $P$ and $Q$ to compute $f\circ
g$.  One runs in linear time, and the other runs in log space, but
neither manages to do both simultaneously.  Do these two programs
represent the same algorithm?

In an application where strict complexity restrictions like linear
time and log space are important, these programs should be considered
different.  But in more relaxed situations, for example if quadratic
time is fast enough, then the difference between these two programs
might be regarded as a mere implementation detail.
\end{ex}

So far, the purposes we have discussed, which may influence the
decision whether two algorithms are the same, have involved
implementing and running the algorithms.  But people do other things
with algorithms, besides implementing and running them.  For someone
who will extend the algorithm to new contexts, or for a mathematician
who will appeal to the algorithm in a proof, the idea underlying the
algorithm is of primary importance.  Furthermore, in both of these
situations the presuppositions of the algorithm will play an important
role.  Thus, for example, it would be an essential characteristic
of (the division form of) the Euclidean algorithm that
it uses a measure of ``size'' on the relevant domain (numerical value
or absolute value in the case of integers, degree in the case of
polynomials) such that one can always\footnote{Or almost always;
division by zero should be excluded.} divide and obtain a remainder
``smaller'' than the divisor.

It is reasonable to suppose that any worthwhile notion of sameness
of algorithms will be adapted to some more or less specific purpose,
and that different purposes will yield different notions of sameness.
A global, all-purpose notion of ``the same algorithm'' is indeed pie
in the sky.

\section{Equivalence}

Is the relation, between programs, of expressing the same algorithm
really an equivalence relation?  The preceding sections suggest that
the relation depends on individual judgment and goals, but suppose we
fix an individual and a goal, and use the resulting judgment; call two
algorithms the same if and only if this particular person regards them
as the same for this particular purpose.  (And don't give him an
opportunity to change his mind.)  Now can we expect to have an
equivalence relation?  Shore's question explicitly asks for an
equivalence relation, but is the intuitive notion of ``the same
algorithm'' necessarily an equivalence relation?

\subsection{Reflexivity}
It seems obvious that every program defines the same algorithm as
itself, that is, that no program defines two algorithms.  But in fact, we
had to take some precautions to ensure this obvious fact.  Recall the
example of the division form and the subtraction form of the Euclidean
algorithm which, when programmed on a Turing machine with unary
notation, can become the same program.  One of our reasons for not
adopting Turing machines as our standard programs was precisely this
possible failure of reflexivity.

Our insistence that ASMs be understood as defining only sequential
algorithms and in fact as defining only their states, initial states,
and transition functions, not the details of what happens within a
step, is also related to reflexivity.  That is because the semantics
of ASMs is defined only at the level of sequential algorithms.  There
is no guarantee that an ASM program cannot be regarded, by two
readers, as describing two different processes within a step; the only
guarantee is that the ultimate outcome of the step must be the same.

We believe that we have taken enough precautions to ensure
reflexivity.  In fact, our discussion of behavioral equivalence in
Section~\ref{beh-eq} was intended to support the thesis that an ASM
program defines the same algorithm as any behaviorally equivalent ASM,
in particular itself.

\subsection{Symmetry}
The colloquial usage of ``the same algorithm'' seems to admit some
mild failures of symmetry.  For example, given a program $A$ at a
rather abstract level and a detailed implementation $B$ of it, a
person reading the text of $B$ and suddenly understanding what is
really going on amid the details might express his understanding by
exclaiming ``Oh, $B$ is just the same as $A$'', while the symmetric
claim ``$A$ is just the same as $B$'' is unlikely.

Nevertheless, the intuitive notion of ``the same algorithm''
underlying Shore's question is clearly a symmetric one.  Shore's
description of the intended intuition refers to the \textbf{unordered}
pair of programs, using the phrases ``two particular programs'' and
``two algorithms'' with no distinction between a first and a second.

Furthermore, someone exclaiming, in the situation described above,
``$B$ is just the same as $A$,'' would probably, if pressed, concede
that he didn't really mean that they are ``the same'' but that $B$ is
merely an implementation of $A$, i.e., that the idea behind $B$ is the
same as that behind $A$, but $B$ contains more details and thereby
hides the idea.

\subsection{Transitivity}
In view of the preceding brief discussion of reflexivity and symmetry,
we regard the question ``Do we have an equivalence relation?'' as
coming down to ``Do we have transitivity?'', and here things are
considerably less clear.  Might there be, for example, finite (but
long) sequences of programs in which each program is essentially the
same as its neighbors (and so should express the same algorithm), yet
the first and last programs would not be considered the same
algorithm?  Here are some examples to indicate the sorts of things
that can happen.  The first example is a continuation of the binary
decision tree example.

\begin{ex}   \label{dec-tree3}
Consider the programs $\Pi_\emp$ and $\Pi'$ described in
Example~\ref{dec-tree}, and modify them, as described earlier, to make
explicit the differences in the work they do within a step.  For
example, have them update Boolean variables $V_s$ with the meaning
``$U_s$ was used.''  These modifications, call them $\widetilde\Pi_\emp$
and $\widetilde\Pi'$, can reasonably be regarded as different algorithms.

There is, however a ``continuous'' (in a discrete sense) transition
between them.  An intermediate stage of this transition would be an
algorithm that begins by looking up the values of $U_s$ for the first
$k$ sequences $s$ (in order of increasing length and lexicographically
for equal lengths, i.e., for the first few levels of the decision tree
and for some initial segment of the next level).  Using the answers,
the algorithm finds (by a single parallel block of guarded updates,
i.e., by a table look-up in a table of size $2^k$) the path through
the part of the tree about which it asked.  It reaches a node $s$ for
which it knows the value of $U_s$ but didn't look up $U_t$ for the
children $t$ of $s$.  The value of $U_s$ tells the algorithm which of
the two children $t$ of $s$ is relevant, so it looks up the value of
that $U_t$.  This value tells which child of $t$ is relevant, so that
determines the next $U$ to evaluate, and so forth.  (The algorithm
also updates the $V$'s as above, to indicate which $U$'s it
evaluated.)  When $k=1$, this algorithm is $\widetilde\Pi_\emp$. When
$k=63$, it is $\widetilde\Pi'$.  These look different.  But is there a
significant difference between the algorithms obtained for two
consecutive values of $k$?  It looks like just a minor bookkeeping
difference.  So we have a sorites situation, going from one algorithm
to an arguably different one in imperceptible steps---imperceptible in
the sense that one could reasonably consider the $k$ and $k+1$
versions mere variant implementations of the same algorithm.
\end{ex}

\begin{ex}
  Similarly, in Example~\ref{pivots}, the quicksort algorithm could
  reasonably be considered unchanged by a minor change in the number
  $k$ of elements whose median determines the first pivot.  But a
  succession of such minor changes can lead to a situation where $k$
  becomes as large as the entire array to be sorted.  Then finding the
  median, to serve as the first pivot, is practically as hard as
  sorting the whole array.  So at this stage, the algorithm is much
  less likely to be regarded as the same as traditional quicksort.
\end{ex}

The next example involves what may be, at least for logicians, the
archetypal example of a trivial syntactic change that makes no real
difference, namely renaming bound variables (also called
$\alpha$-equivalence).  The particular choice of bound variables is so
unimportant that they are often eliminated altogether, for example by
means of de~Bruijn indices in the $\lambda$-calculus \cite{bruijn} or
by the boxes-and-links scheme of Bourbaki \cite{bour}.

\begin{ex}   \label{rename}
Sequential non-interactive ASMs, which we have chosen as our standard
programs, don't have bound variables.  They can, however, and often do
have variables that resemble bound variables in that the algorithm
shouldn't care what those variables are (as long as they are distinct
and don't clash with other variables).  For example, many sequential
ASMs have a dynamic symbol \ttt{mode}, serving as a sort of program
counter, whose value in every initial state is a particular element
(named by) \ttt{start} and whose value at the end of every computation
is \ttt{final}.  Neither the name \ttt{mode} nor the names of its
intermediate values make any real difference to the computation.
Nevertheless, changing these names produces a different ASM, one not
behaviorally equivalent to the original because it has a different set
of states (for a different vocabulary).  At first sight, it seems
clear that two programs that differ only by such a renaming should be
equivalent; the underlying algorithm is the same.

But for certain purposes, there may be a real difference, especially
if the identifiers become very long strings, for example the entire
text of \emph{War and Peace}.  Then an actual implementation will involve
reading the identifiers and (at least) checking which of them are the
same and which are not.  If the identifiers get too long, a real-world
interpreter or compiler will crash.  (For such reasons, compiler
expert Gerhard Goos likes to say that every compiler is almost always
wrong, meaning that it is correct on only finitely many programs.)
Even if the system (interpreter or compiler) is willing to handle very
long identifiers, it must do additional work to distinguish them.
Suppose, for example, that we take a program written with normal-sized
identifiers and modify it by attaching a particular long random string
to the beginning of all the identifiers.  Then, if the compiler reads
from left to right, it will have to read and compare long strings just
to tell whether two occurrences are of the same identifier.  (We could
also attach another long random string, of possibly different length
than the first, to the end of each identifier, so as to defeat
compilers that don't just read from left to right.)

In accordance with the discussion in Section~\ref{beh-eq}, if we want
to take into account the work involved in reading and comparing
identifiers, then we should write the ASM in a way that makes this
work explicit.  Suppose then that this rewriting has been done.  Now
the example leads to a sorites paradox.  Lengthen the identifiers
gradually; at what point do you get a different algorithm?  For a
fixed compiler (or interpreter), there may be a well-defined boundary,
but the boundary moves when the compiler changes.
\end{ex}

Finally, here is an example arising from how the first author, after
grading exam papers, sorts them into alphabetical order for recording
the grades and returning the papers to the class; see also
\cite{merge}.

\begin{ex}   \label{sorting}
With $N$ exams, first find the power of 2, say $2^k$, such that $s =
N/(2^k)$ is at least 3 but less than 6. (We don't have classes with
fewer than 3 students, so $k$ is a non-negative integer.)  Partition
the given, randomly ordered stack of exams into $2^k$ substacks of
size $s$ (rounded up to an integer in some stacks and down in others),
and sort each substack by insertion sort (i.e., by inspection, since
the stacks are so small).  Merge these stacks in pairs to produce
$2^{k-1}$ larger, sorted stacks.  Then merge these in pairs, and so
forth, until all the papers are in one sorted stack.  (Implementation
detail: If desk space is limited, then don't do all the insertion
sorts first and then all the merging.  Rather, the stacks that would
eventually be merged by the algorithm above should be actually merged
as soon as they become available.  This keeps the number of sorted
stacks bounded by $k$ instead of by $2^k$.  Although it's irrelevant
to the purpose of this example, the variation raises again the
question whether it changes the algorithm.)  The decision to have $s$
between 3 and 6 is just a matter of convenience, so that insertion
sort can be easily applied to the stacks of size $s$.  In principle,
the bounds could be any $s$ and $2s$.  For a fixed $N$, varying $s$
gives a chain of algorithms connecting insertion sort (when $2s > N$)
and mergesort (when $s=1$), such that each two consecutive algorithms
in the chain are intuitively, at least for some people's intuition,
not really different algorithms.

If $N$ is not fixed, then there is an infinite sequence of algorithms
as above, indexed by the natural numbers $s$.  But the sequence, which
starts at mergesort, doesn't reach insertion sort.  Instead, there is
another sequence, indexed by increasing $k$ (in the notation above),
that starts at insertion sort (when $k=0$) and heads toward mergesort.
Between the two sequences, there are other algorithms (or other
versions of the same algorithm), for example one that adjusts $k$ (and
therefore $s$) to do half of the work by insertion and half by
merging. (It may be useful to think of this chain of algorithms in a
non-standard model of arithmetic, where $N$ can be infinite, and we
have an internally finite but really infinite chain of algorithms,
indexed by $k$, connecting insertion sort to mergesort.)
\end{ex}

\begin{ex}
Similar comments apply to the usual implementation of quicksort,
where, as described in \cite[page~116]{knuth}, ``Subfiles of $M$ or
fewer elements are left unsorted until the very end of the procedure;
then a single pass of straight insertion is used to produce the final
ordering. Here $M \geq 1$ is a parameter \dots.'' A small change in
the parameter $M$ can be viewed as a mere implementation detail, not a
change in the algorithm, but a succession of such changes can lead,
for files of a fixed size, from quicksort to insertion sort.  The
situation is just like that for mergesort except that insertion sort
is used at the end of quicksort and at the beginning of mergesort.
\end{ex}

In all the preceding examples, we had a finite sequence of programs
such that a reasonable person might call each particular consecutive
pair of algorithms equivalent but might not want to call the first and
last programs of the sequence equivalent.  There is no clear
transition from equivalent to inequivalent, just as in the classical
sorites paradox there is no clear transition from a pile of sand to a
non-pile (or from a non-bald head to a bald one).  That situation
casts doubt on the transitivity of the supposed equivalence relation.

Vop\v enka and his co-workers have developed an interesting approach
to such situations, using the so-called alternative set theory; see
\cite{vopenka}, especially Chapter III.  This theory distinguishes
sets from classes, but the distinction is not just a matter of size as
in the traditonal von~Neumann-Bernays-G\"odel or Kelley-Morse theories
of sets and classes.  Rather, classes are to be viewed as having
somewhat more vague membership conditions than sets do.  As a result,
a subclass of a set need not be a set.

The axiomatic basis of this theory allows a distinction between
natural numbers (in the usual set-theoretic sense) and finite natural
numbers.  If a set contains 0 and is closed under successor, then it
contains all the natural numbers, but a class with the same properties
might contain only the finite ones.  The latter can be regarded as
formally modeling the notion of a feasible natural number.  If an
equivalence relation is a set (of ordered pairs), then its
transitivity property can be iterated any natural number number of
times, but if it is a class of ordered pairs, then only a feasible
number of iterations will work.  Thus, sorites paradoxes disappear if
one accepts that feasibly many of grains of sand do not constitute a
pile and that a man with feasibly many hairs is bald.  These ideas can
be similarly applied to our examples above.  (We thereby add credence
to the first author's claim that the number of exams he has to grade
is not feasible.)

\section{Analogies}

The issues involved in defining equivalence of algorithms are similar
to issues arising in some other contexts, and we briefly discuss a few
of those here.

\subsection{Clustering}

The basic problem of saying when two things (in our case, programs)
should be treated as the same is very similar to the basic problem of
clustering theory, namely saying which elements of a given set are
naturally grouped together.  Indeed, if we had a reasonable metric on
the set  of programs, then we might consider applying the
techniques of clustering theory to it.  But before doing so, we should
take into account a comment of Papadimitriou \cite{papa}:
\begin{quote}
There are
far too many criteria for the `goodness' of a clustering \dots\ and
far too little guidance about choosing among them\dots. The
criterion for choosing the best clustering scheme cannot be determined
unless the decision-making framework that drives it is made
explicit.
\end{quote}
And that brings us back to our earlier comment that the
notion of ``same algorithm'' that one wants to use will depend on what
one intends to do with it and with the algorithms.

We should also note that Papadimitriou's discussion is in the context
where the metric is given and the issue is to choose a criterion for
goodness of clustering.  Our situation is worse, in that we do not
have a natural metric on the set of programs.

\subsection{Proofs}

Another analog of ``When are two algorithms the same?'' is ``When are
two proofs the same?''  The role played by the set of programs in our
discussion of algorithms would be played by the set of formal
deductions in some axiomatic system.  The problem is to say when two
formal deductions represent the same proof.  There are many axiomatic
systems that one might consider, ranging from propositional calculus
up to systems like Zermelo-Fraenkel set theory that provide a
foundation for almost all of mathematics.  But the question of which
deductions represent the same proof makes sense at each level, and it
leads to many of the same issues that we have discussed in connection
with algorithms.

Equality of proofs has been studied in considerable detail but usually
only at the level of elementary logic; see for example \cite{lambek}
and \cite{girard}.  Renaming bound variables is regarded as not
changing a proof; the same goes for cancelling the introduction of a
connective or quantifier and an immediately following elimination of
that connective or quantifier.  (This formulation refers to the
introduction and elimination rules of natural deduction systems, but
there are analogous notions for sequent calculi.  Hilbert-style
calculi are not used for studies of equality of proofs, because they
require circumlocutions that make formal deductions bear little
resemblance to the intuitive proofs they should express.)  But such
equivalence relations between deductions are too fine to capture the
intuitive notion of proofs being the same.  In this respect, they
resemble the equivalence relation described in \cite{yanof}.  In fact
there is another similarity here, namely that in both cases the
equivalence relations are designed to look nice from a
category-theoretic viewpoint.

Considerably larger rearrangements of the material in a deduction
would be recognized by mathematicians as not changing the proof that
the deduction represents.  Other changes, however, are considered
essential, for example the difference between an analytic and an
elementary proof in number theory, or the difference between a
bijective proof and a manipulation of generating functions in
combinatorics.  Between the ``clearly the same'' and ``clearly
different'' cases, there is a gray area that looks quite analogous to
what we see in the case of algorithms.

The analogy between proofs and algorithms has been given mathematical
content by the propositions-as-types or Curry-Howard correspondence.
Here formal deductions in certain systems correspond exactly to
programs in certain (usually functional) programming languages.  So
there may be reason to think that, if we thoroughly understood
sameness in one of the two contexts, then we could transfer that
understanding (or at least part of it) to the other.  Unfortunately,
the existing notions of equivalence of proofs translate into
equivalence relations on algorithms that are too fine; they count
algorithms as equivalent only if they are so for trivial reasons.

In addition to the Curry-Howard correspondence, there is another
connection between proofs and computation.  If we take an algorithm
for computing a function $f$ and we run it on input $x$ obtaining
output $y$, then the record of the computation can be regarded as a
proof, in an appropriate formal system, of the equation $f(x)=y$.
(This is the essential point behind the theorem that all recursive
functions on the natural numbers are representable in (very weak)
systems of arithmetic like Robinson's $Q$; for details see
\cite[Section~2.4]{tmr} and note that what is nowadays called
representable was there called definable.)  If we had a good notion of
equivalence of proofs in such formal systems, then we could use it to
define a notion of equivalence of algorithms.  Call two algorithms,
for computing the same function $f$, equivalent if, for each input
$x$, not only do they produce the same output $y$, but the resulting
proofs of $f(x)=y$ are equivalent.  Unfortunately, it seems that, just
as in the case of the Curry-Howard correspondence, existing notions of
equivalence of proofs are too fine and therefore so are the resulting
notions of equivalence of algorithms.

\subsection{Other analogous questions}

We already mentioned, in Remark~\ref{ideal}, that certain algorithms
are considered different because they are based on different ideas.
So we ask: What exactly does it mean for two ideas to be different?
What is the ``right'' equivalence relation on ideas?  This problem
looks even worse than the original question about algorithms, partly
because of its great vagueness and partly because the only analog for
the set of programs, a set of things that can express ideas, would
seem to be something like the set of all (English?) texts, and the
ways in which a text can express an idea seem to be entirely out of
control.
A related question about texts is
``What does it mean for a text in one language to be a translation of
a text in another language?''  Is expressing the same idea a necessary
condition?  Is it a sufficient condition?  Does it depend on the
chosen notion of ``same idea'''?

Moschovakis \cite{ynm} has argued that the meaning of a term in
English (or other natural languages) is the algorithm for computing
its denotation.  With this identification of meanings with algorithms,
the question of when two ideas are the same would be not merely
analogous to the question of when two algorithms are the same; the
former would become a special case of the latter, at least for those
ideas that occur as the meanings of terms.

Similar issues arise in mathematics.  What does it mean to say that
two theorems are equivalent?  Material equivalence, meaning equality
of truth values, is certainly not the intention.  The intuitive idea
is that each of the theorems follows easily from the other -- at least
more easily than proving either theorem from scratch.  But ``easy''
and ``from scratch'' are subjective notions.

Reverse mathematics (see \cite{hf,sosoa}) provides a notion of
equivalence of theorems, namely provable equivalence over some weak
base theory. This notion is, however, much coarser than the intuitive
notion, since the proofs of equivalence may be highly non-trivial,
often more difficult (because of the weakness of the base theory) than
the usual proofs (in strong theories like Zermelo-Fraenkel set theory)
of the theorems themselves.  Indeed, much of the fascination of reverse
mathematics comes from the equivalence between theorems from very
different fields of mathematics, theorems that at first seem to have
nothing to do with each other.  Another way to view the coarseness of
the notion of equivalence given by reverse mathematics is that it
looks only at one aspect of theorems, namely the set-existence
assumptions that underlie them, whereas the intuitive notion of
equivalence would look rather at the (intuitive) content of the
theorems.

We trust that the reader can extend this list of examples.

\end{document}